\PassOptionsToPackage{unicode}{hyperref}
\PassOptionsToPackage{hyphens}{url}
\documentclass[12pt]{article}
\usepackage{amsmath}
\usepackage{graphicx,psfrag,epsf}
\usepackage{enumerate}
\usepackage[]{natbib}
\usepackage{textcomp}

\newcommand{\blind}{0}

\providecommand{\tightlist}{%
  \setlength{\itemsep}{0pt}\setlength{\parskip}{0pt}}

\usepackage{longtable,booktabs,array}
\usepackage{calc} % for calculating minipage widths
\usepackage{etoolbox}
\makeatletter
\patchcmd\longtable{\par}{\if@noskipsec\mbox{}\fi\par}{}{}
\makeatother
\IfFileExists{footnotehyper.sty}{\usepackage{footnotehyper}}{\usepackage{footnote}}
\makesavenoteenv{longtable}

\usepackage{booktabs}
\usepackage{longtable}
\usepackage{array}
\usepackage{multirow}
\usepackage{wrapfig}
\usepackage{float}
\usepackage{colortbl}
\usepackage{pdflscape}
\usepackage{tabu}
\usepackage{threeparttable}
\usepackage{threeparttablex}
\usepackage[normalem]{ulem}
\usepackage{makecell}
\usepackage{xcolor}

\IfFileExists{bookmark.sty}{\usepackage{bookmark}}{\usepackage{hyperref}}
\IfFileExists{xurl.sty}{\usepackage{xurl}}{} % add URL line breaks if available
\hypersetup{
  pdftitle={Evidencing preferential attachment in dependency network evolution},
  pdfkeywords={R packages, Poisson regression, Bayesian hierarchical model, degree distribution, scale-free},
  hidelinks,
  pdfcreator={LaTeX via pandoc}}

\begin{document}

\def\spacingset#1{\renewcommand{\baselinestretch}%
{#1}\small\normalsize} \spacingset{1}

%%%%%%%%%%%%%%%%%%%%%%%%%%%%%%%%%%%%%%%%%%%%%%%%%%%%%%%%%%%%%%%%%%%%%%%%%%%%%%

\if0\blind
{
  \title{\bf Evidencing preferential attachment in dependency network evolution}

  \author{
        Clement Lee \\
    School of Mathematics, Statistics and Physics, Newcastle University, UK\\
      }
  \maketitle
} \fi

\if1\blind
{
  \bigskip
  \bigskip
  \bigskip
  \begin{center}
    {\LARGE\bf Evidencing preferential attachment in dependency network evolution}
  \end{center}
  \medskip
} \fi

\bigskip
\begin{abstract}
Preferential attachment is often suggested to be the underlying mechanism of the growth of a network, largely due to that many real networks are, to a certain extent, scale-free. However, such attribution is usually made under debatable practices of determining scale-freeness and when only snapshots of the degree distribution are observed. In the presence of the evolution history of the network, modelling the increments of the evolution allows us to measure preferential attachment directly. Therefore, we propose a generalised linear model for such purpose, where the in-degrees and their increments are the covariate and response, respectively. Not only are the parameters that describe the preferential attachment directly incorporated, they also ensure that the tail heaviness of the asymptotic degree distribution is realistic. The Bayesian approach to inference enables the hierarchical version of the model to be implemented naturally. The application to the dependency network of R packages reveals subtly different behaviours between new dependencies by new and existing packages, and between addition and removal of dependencies.
\end{abstract}

\noindent%

\vfill

\newpage
\spacingset{1.9} % DON'T change the spacing!

\hypertarget{intro}{%
\section{Introduction}\label{intro}}

Preferential attachment (PA) is an important concept in network science that describes how networks, such as social networks, citation networks, and the internet, grow and evolve over time. While the precise mechanics vary in the literature, the common rule is that, as new vertices join the network and new edges are created over time, the probability of an existing vertex getting further connected is proportional to a preference function \(g(k)\), where \(k\) is the total degree or in-degree (collectively termed as the ``degree'') of the vertex. To achieve the preferential effect, \(g(k)\) is usually non-decreasing with \(k\). Under the vanilla model by \citet{ba99a}, one new vertex joins the network at a time and connects with one existing vertex, with \(g(k)=k\). While this model, called linear PA hereafter, is itself an embodiment of PA in terms of citations, similar ideas had been proposed by \citet{yule25}, \citet{simon55} and \citet{price76}.

Closely related to PA is the notion of scale-freeness, or equivalently degree distributions following a power law, which is defined as \(f(k)\propto{}k^{-\gamma}\), where \(f\) is the probability mass function, and \(\gamma>1\) the exponent. While the definition of scale-free has been relaxed by e.g.~\citet{cf19} and \citet{vvvk19}, we use it in the strict sense for illustration. Subsequently, \(\bar{F}(k)\propto{}k^{-(\gamma-1)}\), where \(\bar{F}\) is the survival function, and \(\log{}f(k)\) and \(\log{}\bar{F}(k)\) are both linear with \(\log{}k\), albeit with different slopes.

Based on this relationship, many real networks have their empirical degree distribution or survival function deemed close to a straight line on the log-log scale, and are therefore considered scale-free; see Table~1 of \citet{dm02}. On the theoretical side, the asymptotic degree distribution under linear PA is shown by \citet{brst01} and \citet{bbcr03} to follow a power law. This match between theoretical properties and empirical observations opens up the possibility of attributing the growth of a network to PA, simply by analysing its empirical degree distribution at one or more snapshots.

There are, however, a few issues if one attempts to equate scale-freeness with PA through the lens of the degree distribution. First, a general \(g(k)\) does not necessarily lead to scale-freeness. \citet{krl00} and \citet{kr01} considered the power preference function \(g(k)=k^{\alpha}\) \((\alpha>0)\), and found that the degree distribution is asymptotically Weibull when \(0<\alpha<1\), and a finite number of vertices receives nearly all edges when \(\alpha>1\). Second, while (a special case of) PA implies that the network is scale-free, the converse is not true. For example, \citet{hofstad16} showed that both the generalised random graph and the configuration model imply scale-freeness. \citet{hbb20} and \citet{hb23} proposed models in which the resulting scale-free networks differ to those generated by PA in terms of network robustness. The absence of such causality also suggests that degree-distribution-based hypothesis tests \citep{mohdzaid16} are insufficient for deciding if a real network comes from PA.

Third, as a result of the ambiguity on the definition of ``being close'' to a power law, there has been a controversy on whether scale-freeness is ubiquitous in real networks. \citet{bc19} argued that scale-free networks are rare in reality by testing the power law against commonly used alternatives. Their claim was rebutted by \citet{asvw20} and \citet{vvvk19}, with the latter considering a network scale-free if its empiricial degree distribution is heavy-tailed\footnote{What we call heavy-tailed here is what \citet{vvvk19} called \emph{regularly varying}.} in extreme value theory (EVT) terminology. \citet{lef24} used a different formulation based on EVT to reveal that many networks are \emph{partially} scale-free, meaning that the body of the empirical degree distribution follows a power law, thus echoing \citet{dm02} who pointed out that the linearity on the log-log plot usually takes place within a rather narrow range. Moreover, its tail deviates from the power law but is still heavy. Not only does such subtlety imply the inadequateness of the alternative distributions used by \citet{bc19} as they are all light-tailed, it also highlights a peculiarity that the fit for the largest degrees, which correspond to the most influential players in the network, is often sacrificed for that for the smaller degrees.

\begin{figure}[!tp]

{\centering \includegraphics[width=0.48\linewidth]{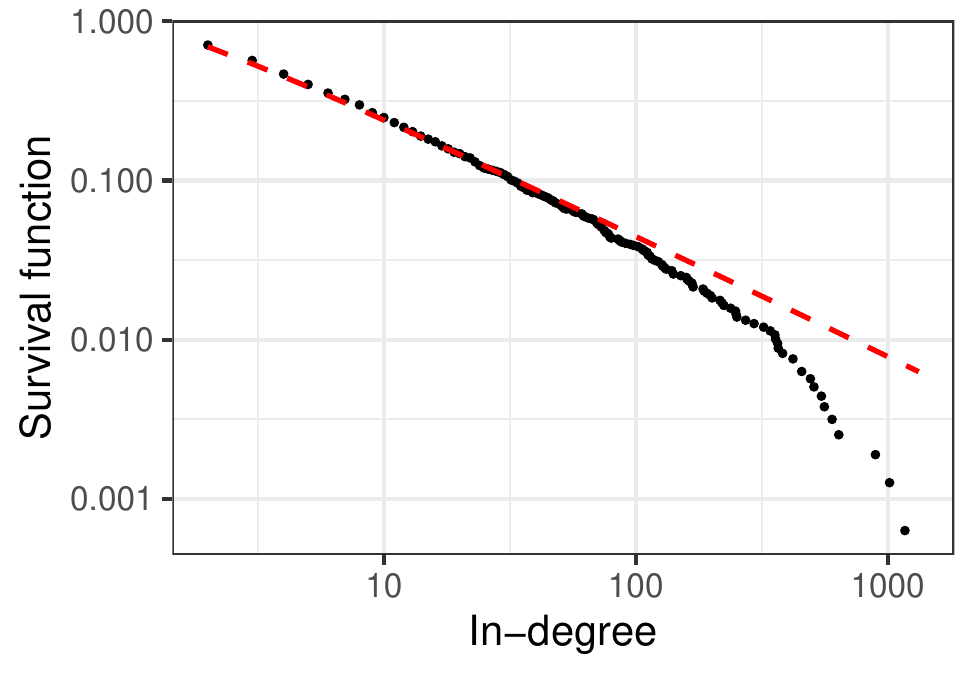} 

}

\caption{Empirical survival function (dots) of \texttt{Imports} between CRAN packages on 2019-01-29 on the log-log scale. The dashed line is the fitted line according to the power law.}\label{fig:cran-deg}
\end{figure}

We motivate our work by illustrating the partial scale-freeness of the dependency network of R packages on the Comprehensive R Archive Network (CRAN)\footnote{cran.r-project.org}, the data set analysed in this paper. The empirical survival function of the in-degrees of the packages, for one type of dependencies on one day, is plotted in Figure~\ref{fig:cran-deg} on the log-log scale. The dashed line, which is fitted according to the power law, shows the deviation of the largest in-degrees. The same phenomenon of partial scale-freeness is observed throughout the data collection period and across different types of dependencies.

Summarising the above issues, a general PA model does not imply scale-freeness, scale-freeness does not imply PA as the generating mechanism, and many real networks are at best partially scale-free. Note that these issues arise mainly because the interest is in the degree distribution, which is nonetheless a useful summary when only one or a few snapshots of the network are available. However, taking a step back, if data on the evolution of the network is available, \emph{can we directly measure PA by evidencing the rule stated at the beginning of the paper, while taking into account the findings on scale-freeness?}

There have been efforts on measuring PA in the presence of evolution data. \citet{jnb03} used a differential equation approach and focused on the cumulative function of the \emph{normalised} power preference function, to find linear PA for two networks and sublinear PA for another two. \citet{newman01d} and \citet{pss16} both estimated the preference function, with the former using a nonparametric approach to compute what they called the relative probability of a vertex of a given degree receiving a new edge, and the latter incorporating fitness in their Bayesian parametric approach. By measuring on the macroscopic scale, these works circumvented the issue of the increments of the evolution fluctuating at finer time intervals.

Recent works properly treated these increments as observations of random variables in a statistical model. \citet{obu19} proposed a generalised linear model (GLM) approach where the binary response is the choice made by a vertex to connect with another, while the covariates are measures of various network properties such as the degree, thus incorporating PA naturally. \citet{gp22} proposed a close alternative termed the repeat-choice model and, similar to \citet{obu19}, sampled edges from the whole of data in their applications in order to reduce the computational burden. \citet{inoue22} incorporated transitivity in a way similar to \citet{obu19} did fitness, and extended their model to hypergraphs. \citet{arnold21} and \citet{amc21} proposed a mixture preference function in their detection of how PA and uniform attachment change over time. When dealing with the issue of multiple new vertices and edges at each time point, which is common in network evolution data, they used a sampling procedure to approximate the likelihood instead of calculating the exact probabilities of weighted sampling without replacement, which arise naturally when adhering to the rule of PA. Such computational issue could be resolved by the latent variable approach of \citet{lo23}, who however mainly focused on when evolution data is not available.

In this paper, we follow the GLM approach for modelling the increments of the network evolution on the most granular level. With the use of a distribution property, the common formalisation of PA lends itself to a Poisson regression model, where the response is the increment of the degree and the covariate is the degree itself. Not only does this lead to a quick diagnostic of PA based on the direct relationship between the \emph{expected} increment and the degree, it also enables efficient and exact computations of the likelihood, without the need of ad-hoc sampling procedures. As the preference function is naturally incorporated in the model, its parameters can be inferred directly. Furthermore, in addition to the popular yet somewhat problematic power preference function, we consider a preference function that implies partial scale-freeness, thus appropriate for real networks with regard to the tail of the empirical degree distribution. This \emph{tail-realistic} preference function will be shown to clearly outperform the power preference function in some of our applications. Lastly, the parameters are allowed to vary over time and inferred under a Bayesian hierarchical version of the model.

The rest of the article is as follows: Section~\ref{preliminaries} provides the preliminaries and devises a diagnostic for PA, which is used in the exploratory data analysis of the CRAN dependencies data in Section~\ref{eda}. Section~\ref{model} introduces the full model for the increments of network evolution, and Section~\ref{inference} provides the inference procedure. The hierarchical version of the model is introduced in Section~\ref{hierarchical}. Section~\ref{application} shows the results of fitting the model to the CRAN dependencies data, and Section~\ref{discussion} concludes the paper.

\hypertarget{preliminaries}{%
\section{Preliminaries}\label{preliminaries}}

In this section, we introduce the notation used hereafter and some model assumptions, in order to arrive at a simple diagnostic tool for PA, which will be used in Section~\ref{eda}.

Consider the evolving network at time \(t\geq0\) to be represented by an unweighted directed graph \(\mathcal{G}_t:=(\mathcal{V}_t,\mathcal{E}_t)\), where \(\mathcal{V}_t\) and \(\mathcal{E}_t\) are the vertex set and edge set, respectively. At time \(t(>0)\), we assume that \(M_t\geq0\) new vertices join and never leave the network. This means \(\{\mathcal{V}_t\}_{t\geq0}\) is a filtration, and the number of vertices at time \(t\) is \(n_t:=|\mathcal{V}_t|=|\mathcal{V}_{t-1}|+M_t=|\mathcal{V}_0|+\sum_{s=1}^tM_s\). When the changes from time \(t-1\) to \(t\) are concerned, those in \(\mathcal{V}_{t-1}\) are referred to as \emph{existing} vertices.

Regarding the changes in the edge set, we assume that there are both added and deleted edges at time \(t\), meaning that neither \(\mathcal{E}_{t-1}\) nor \(\mathcal{E}_{t}\) is a subset of each other. The set of new edges, i.e.~\(\mathcal{E}_t\backslash\mathcal{E}_{t-1}\) is partitioned into \emph{internal} and \emph{external} ones, while the deleted edges constitute the set \(\mathcal{E}_{t-1}\backslash{}\mathcal{E}_t\). We can define several vertex-specific quantities regarding the in-degree and increments thereof. For an existing vertex \(i(\in\mathcal{V}_{t-1})\), its \emph{internal} increment at time \(t\) is the number of new edges from existing vertices to \(i\), denoted by \(X_{i,t}:=|\{u:(u,i)\in\mathcal{E}_t\backslash\mathcal{E}_{t-1}\wedge{}u\in\mathcal{V}_{t-1}\}|\), and, collectively, \(\mathbf{X}:=(X_{i,t})_{i\in\mathcal{V}_{t-1},t>0}\). Similarly, the \emph{external} increment is the number of new edges from new vertices to \(i\), denoted by \(Y_{i,t}:=|\{u:(u,i)\in\mathcal{E}_t\backslash\mathcal{E}_{t-1}\wedge{}u\in\mathcal{V}_t\backslash\mathcal{V}_{t-1}\}|\), and \(\mathbf{Y}:=(Y_{i,t})_{i\in\mathcal{V}_{t-1},t>0}\). Next, the \emph{deletion} increment is the number of deleted edges from existing vertices to \(i\), denoted by \(Z_{i,t}:=|\{u:(u,i)\in\mathcal{E}_{t-1}\backslash{}\mathcal{E}_t\}|\), and \(\mathbf{Z}:=(Z_{i,t})_{i\in\mathcal{V}_{t-1},t>0}\). Lastly, the \emph{current} in-degree of vertex \(i\) at time \(t\) is \(k_{i,t}:=|\{u:(u,i)\in\mathcal{E}_t\}|\), and \(\mathbf{k}:=(k_{i,t})_{i\in\mathcal{V}_t,t>0}\). The scalar quantities defined in this paragraph are related by
\begin{align}
k_{i,t}&=k_{i,t-1}+X_{i,t}+Y_{i,t}-Z_{i,t}.
\label{eq:kIJD}
\end{align}
When it comes to data analysis, the internal, external and deletion increments are collectively called the three \emph{categories} of increments.

Now, consider the directed version of linear PA, where at each time point 1 vertex joins the network and brings about 1 directed edge towards an existing vertex. Using our notation, \(\Pr(M_t=1)=1\) for all \(t\), and both \(\mathcal{I}_t\) and \(\mathcal{D}_t\) are empty. This existing vertex is chosen at random with probability proportional to the current in-degree. Used in the original model of \citet{ba99a} was the total degree, which however is simply in-degree plus 1; here we opt for the in-degree. Mathematically, the random variables \(Y_{1,t},Y_{2,t},\ldots,Y_{n_{t-1},t}|M_t(=1)\) jointly follow the multinomial distribution with parameters \(\left(M_t;k_{1,t-1}\left/\sum_{j:j\in\mathcal{V}_{t-1}}k_{j,t-1}\right.,k_{2,t-1}\left/\sum_{j:j\in\mathcal{V}_{t-1}}k_{j,t-1}\right.,\ldots,k_{n_{t-1},t-1}\left/\sum_{j:j\in\mathcal{V}_{t-1}}k_{j,t-1}\right.\right)\). The conditioning on \(M_t\) will become useful in the next paragraph.

We generalise the model by letting the probability of being chosen to be proportional to \(k^{\alpha}\) (\(\alpha>0\)), i.e.~the \emph{power} preference function. This means that the parameters of the multinomial distribution are now \(\left(M_t;\tilde{k}_{1,t-1}^{\alpha},\tilde{k}_{2,t-1}^{\alpha},\ldots,\tilde{k}_{n_{t-1},t-1}^{\alpha}\right)\), where \(\tilde{k}_{i,t}^{\alpha}:=k_{i,t}^{\alpha}\left/\sum_{j:j\in\mathcal{V}_t}k_{j,t}^{\alpha}\right.\) is the \emph{normalised} weight, with the dependence on the in-degrees of other vertices suppressed in the notation. Linear PA is obtained when \(\alpha=1\), while the cases \(\alpha<1\) and \(\alpha>1\) are termed sublinear and superlinear PA, respectively \citep{kk08}. Further generalising, if we assume \(\{M_t\}_{t>0}\) is a sequence of independent and identically distributed (iid) Poisson random variables with mean \(\mu\), we can use the multinomial splitting property \citep{kingman93} to marginalise \(M_t\), and arrive at \emph{independent} Poisson distributions for \((Y_{1,t},Y_{2,t},\ldots,Y_{n_{t-1},t})\) with means \(\left(\mu\tilde{k}_{1,t-1}^\alpha,\mu\tilde{k}_{2,t-1}^\alpha,\ldots,\mu\tilde{k}_{n_{t-1},t-1}^\alpha\right)\). Equivalently,
\begin{align}
Y_{i,t}&\overset{\text{ind}}{\sim}\mu\tilde{k}_{i,t-1}^\alpha,\qquad{}i\in\mathcal{V}_{t-1},t>0.
\label{eq:power}
\end{align}
If we take the logarithm of the expectation of \(Y_{i,t}\), we have
\begin{align}
\log{}E(Y_{i,t})&=\alpha\log{}k_{i,t-1}+\log\mu-\log\left(\sum_{j:j\in\mathcal{V}_{t-1}}k_{j,t-1}^\alpha\right),
\label{eq:log-log}
\end{align}
meaning that, for any existing vertex \(i\), the expected increment and the in-degree are linear on the log-log scale with slope \(\alpha\). Whenever the relationship between these two variables is concerned, it is understood that the in-degree is for the previous time point.

The implication is that, as preliminary evidence of PA for real data, we can plot the actual increments against the in-degrees on the log-log scale to see if the points lie on a straight line. While adjustments are required to accommodate the discrete nature of the actual increments, this provides a simple yet useful visual diagnostic for evidencing PA. Note that this only applies when the power preference function is concerned, and is unrelated to the log-log plot of empirical degree distribution for checking scale-freeness.

\hypertarget{eda}{%
\section{Data and exploratory analysis}\label{eda}}

In this section we look at the data, of which the in-degree distribution of a snapshot is plotted in Figure~\ref{fig:cran-deg}, and carry out the diagnostic proposed in Section~\ref{preliminaries}. We also examine the joint distribution of the in-degrees and out-degrees, to rule out the need to model their potential dependence.

Using the function \texttt{CRAN\_package\_db()} in the R package \texttt{tools} \citep{cran25}, the dependencies of all R packages on CRAN were collected daily from 2019-01-29 to 2024-12-31 inclusive. While there were other ecosystems of R packages such as Bioconductor\footnote{bioconductor.org}, CRAN is the largest of all and provides a natural boundary of the network concerned. As there are some dates where no changes were made (during holiday periods or when maintenance work was carried out), we shall refer to successive dates or time points as those \emph{where there were changes}. Each row of the data set details the names of the packages at the two ends of the dependency, its \emph{type} (\texttt{Imports}, \texttt{Depends}, \texttt{Linking\ to}, \texttt{Suggests}, \texttt{Enhances}), whether the dependency was added or deleted compared to the previous date, and the previous and current dates. While there are some packages without any dependencies of any type on other packages (nor were they depended on by other packages), the number of such packages, which are subsequently excluded from the dependency network, is negligible. On the other hand, within those who have ever joined the network, there are packages which have been deleted from CRAN for not meeting certain requirements. To preserve the filtration assumption of the vertex set, in data analysis we consider their \emph{dependencies} deleted, but not the packages themselves.

\begin{figure}[!ht]

{\centering \includegraphics[width=0.48\linewidth]{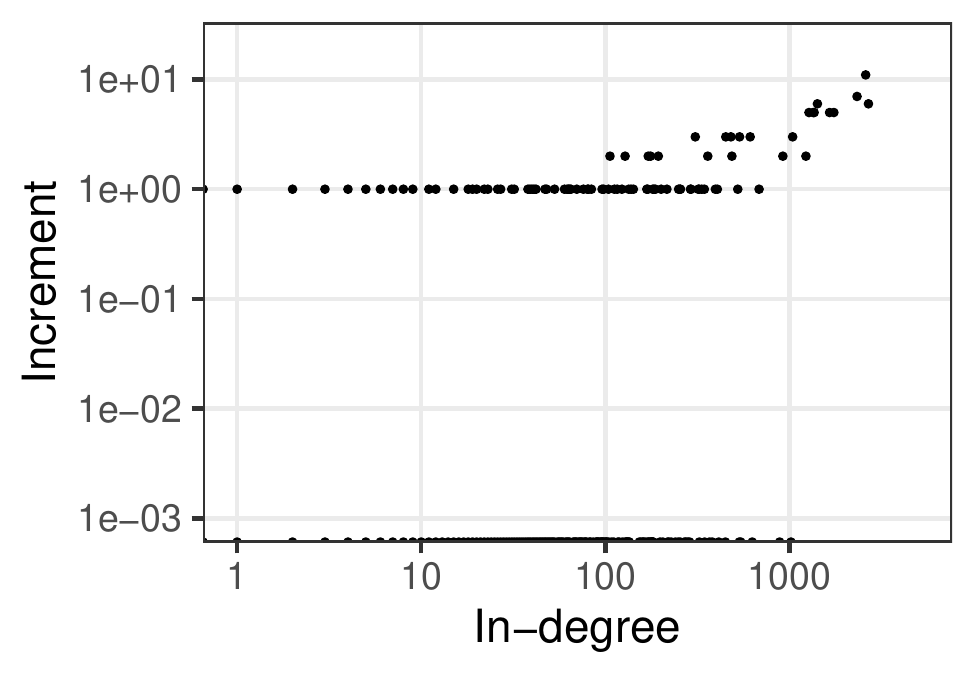} \includegraphics[width=0.48\linewidth]{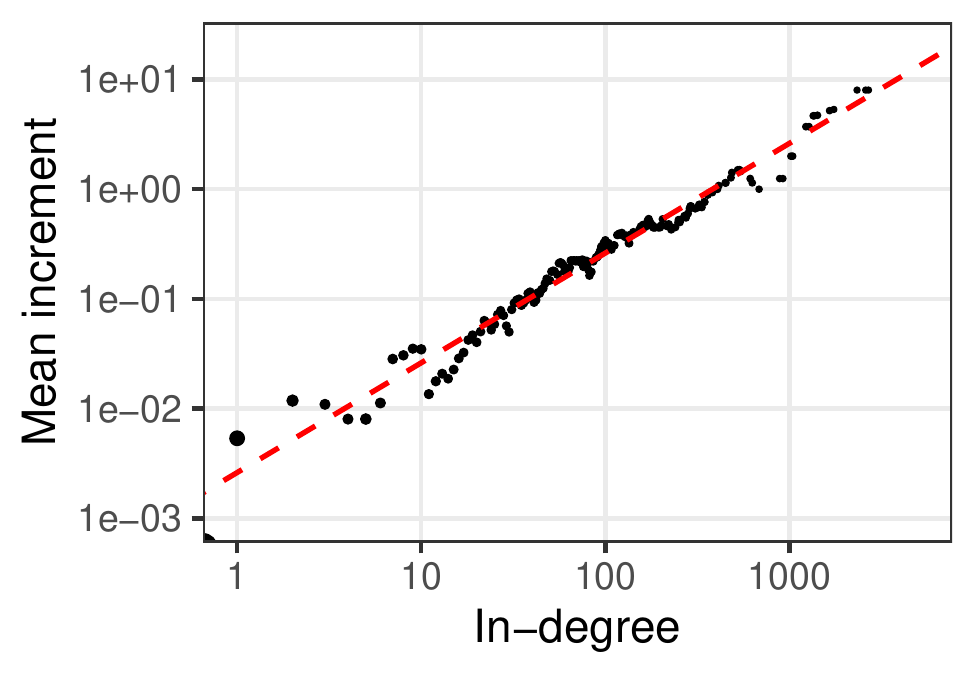} 

}

\caption{Left: External increments against in-degree, of \texttt{Imports} between CRAN packages on 2022-08-22, on the log-log scale. Some points lie on the boundary of the box because 0 cannot be properly included on the log-log scale. Right: Smoothed averages on the same scale and overlaid by the line of best fit with the slope fixed to 1 (dashed).}\label{fig:increments-vs-degrees}
\end{figure}

Upon aggregating the dependencies of the type \texttt{Imports} by the dependend packages, we plot the external increment against the in-degree on the log-log scale on the left of Figure~\ref{fig:increments-vs-degrees}, which however displays an obscure pattern as the increments are discrete and contain mostly \(0\)'s or \(1\)'s. Not only do the smoothed averages on the right suggest that the increment on average increases with the in-degree, but they lie close to a straight line, hence aligning with what Equation~\eqref{eq:log-log} implies. This is the preliminary evidence of PA and, more specifically, PA with a power preference function. However, instead of finding the line of best fit and using its slope as the estimate of \(\alpha\), we overlay the points with the line of best of \emph{with the slope fixed to 1}, indicating that the potential PA is close to a linear one. This is because the tail behaviour of the implied limiting degree distribution is very different under sublinear and superlinear PA \citep{krl00, kr01} and with linear PA, and it is not sensible to base the value of \(\alpha\) on an estimate with substantial uncertainty, as it was obtained by smoothed averages of the increments at a single time point. The power preference function will be revisited in Section~\ref{model}, alongside an alternative preference function that is \emph{tail-realistic}.

\begin{figure}[!h]

{\centering \includegraphics[width=0.48\linewidth]{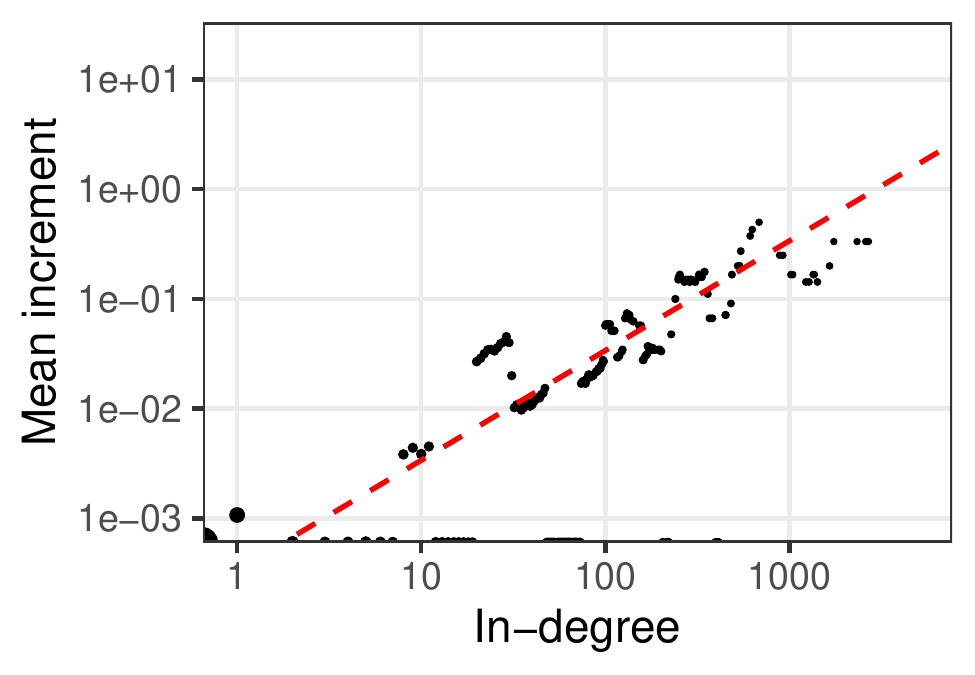} \includegraphics[width=0.48\linewidth]{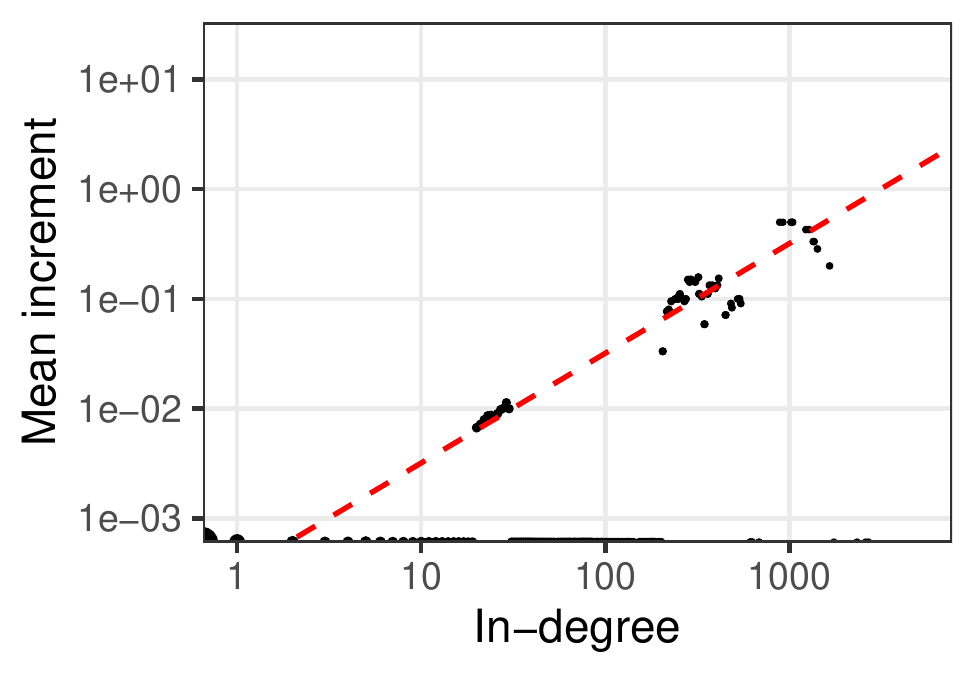} 

}

\caption{Smoothed averages of internal (left) and deletion (right) increments against in-degree, of \texttt{Imports} between CRAN packages on 2022-08-22, on the log-log scale. Some points lie on the boundary of the box because 0 cannot be properly included on the log-log scale. Overlaid is the line of best fit with the slope fixed to 1 (dashed).}\label{fig:internal-deletion}
\end{figure}

The above phenomenon of smoothed averages for a single date being close to a straight line of slope \(1\) on the log-log scale has been observed for the internal and deletion increments in Figure~\ref{fig:internal-deletion} as well as for other dates (not shown). These two categories will therefore be considered in the same way, albeit separately, in the modelling. On the other hand, patterns vary when it comes to other types of dependencies due to their very nature. We shall focus on the hard dependencies only, which do not allow reciprocity, meaning if there is a dependency of package \(A\) on package \(B\) there cannot be a dependency of the same type of \(B\) on \(A\). Within the hard dependencies, \texttt{Imports}, which has the largest share of data, and the union of \texttt{Imports} \(+\) \texttt{Depends}, which together take up more than half of the data, will be considered, but not \texttt{Linking\ to}, as it has a minority share due to its very nature. The soft dependencies (\texttt{Suggests} and \texttt{Enhances}) are not considered here partly because that reciprocity is allowed in the data. The implication is that the in-degrees and out-degrees are potentially correlated, not only in the body of the data but also in the tail, and their joint distribution should be properly modelled using the approach by, for example, \citet{cw24} and \citet{wr24}.

\begin{figure}[!b]

{\centering \includegraphics[width=0.48\linewidth]{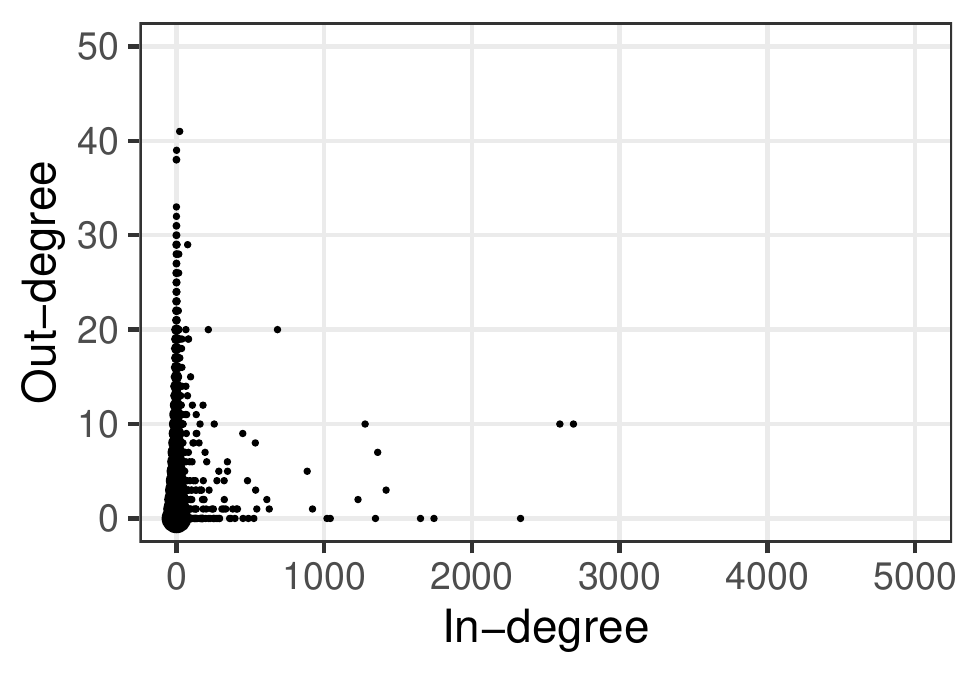} \includegraphics[width=0.48\linewidth]{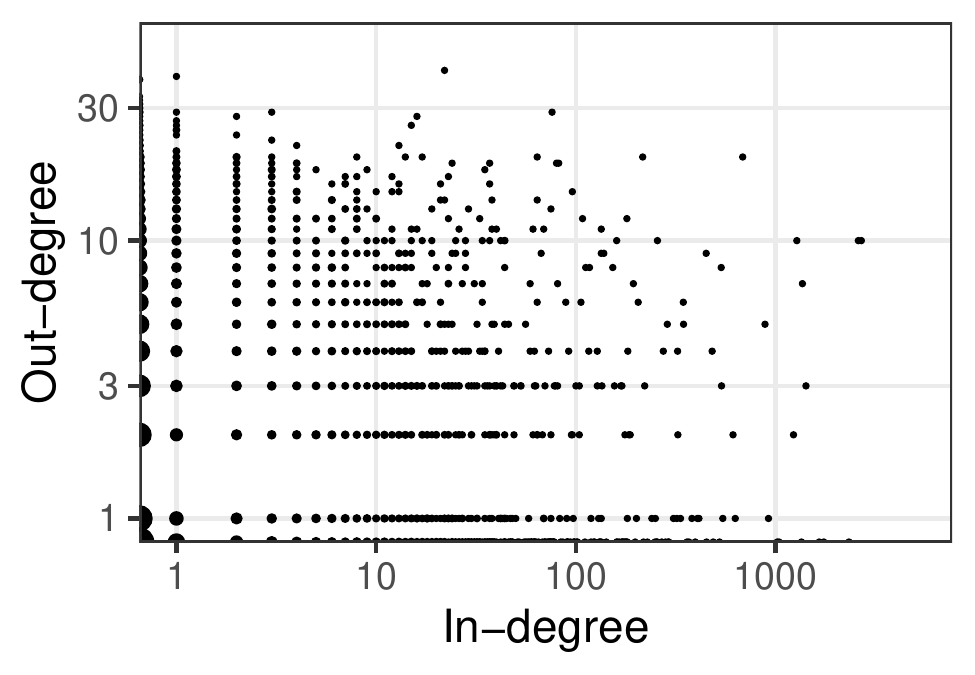} 

}

\caption{Out-degrees against in-degrees, of \texttt{Imports} between CRAN packages on 2022-08-22, on the original (left) and log-log (right) scale.}\label{fig:example-degrees}
\end{figure}

For \texttt{Imports} and \texttt{Imports} \(+\) \texttt{Depends}, there should be no need of joint modelling of the in- and out-degrees due to that reciprocity is not permitted for hard dependencies. Nevertheless, for the same subset of data as in Figure~\ref{fig:increments-vs-degrees}, we plot the out-degrees against the in-degrees in Figure \ref{fig:example-degrees}, which suggests weak correlation both in the body (Pearson correlation \(=-0.009\)) and in the tail (as the two variables are not simultaneously large). This phenomenon applies to other dates and across both types, and so such joint modelling as that by \citet{wr24} will not be incorporated in our model.

\hypertarget{model}{%
\section{Model and likelihood}\label{model}}

In this section, using the external increments \(\mathbf{Y}=(Y_{i,t})_{i\in\mathcal{V}_{t-1},t>0}\) for illustration throughout, we propose the full model for regressing the increments on the in-degrees.

A few assumptions have to be made. First, while \(k_{i,t}\) could be seen as a random variable due to Equation~\eqref{eq:kIJD}, modelling the process of in-degrees is not of interest here. Rather, they are viewed as covariate hence given, with the increments being the responses. Second, and relatedly, conditional on \((k_{i,t-1})_{i:i\in\mathcal{V}_{t-1}}\), \((X_{i,t})_{i:i\in\mathcal{V}_{t-1}}\), \((Y_{i,t})_{i:i\in\mathcal{V}_{t-1}}\) and \((Z_{i,t})_{i:i\in\mathcal{V}_{t-1}}\) are modelled separately, as the focus is on the in-degree-increment relationship. Third, these three increments are assumed independent of their values at all previous time points. Essentially, the sequences of increments have the first-order Markov property.

If we generalise the preference function of vertex \(i\) at time \(t\) from \(k_{i,t}^{\alpha}\) to a general \(g(k_{i,t}|\boldsymbol{\theta})\), where \(\boldsymbol{\theta}\) is the parameter vector, then the model distribution assumption in Equation~\eqref{eq:power} becomes
\begin{align}
Y_{i,t}&\overset{\text{ind}}{\sim}\mu\tilde{g}(k_{i,t-1}|\boldsymbol{\theta}),\qquad{}i\in\mathcal{V}_{t-1},t>0,
\label{eq:dist}
\end{align}
where, similar to before, \(\tilde{g}(k_{i,t}|\boldsymbol{\theta}):=g(k_{i,t}|\boldsymbol{\theta})\left/\sum_{j:j\in\mathcal{V}_t}g(k_{j,t}|\boldsymbol{\theta})\right.\) is the normalised weight. Note that \(\tilde{g}(k_{i,t-1}|\boldsymbol{\theta})\) is influenced by \(\{k_{j,t-1}\}_{j\in\mathcal{V}_{t-1}}\), which is not explicit because of the notational suppression. However, as we treat all of the in-degrees, the preference function and the parameters as given, the independence assumption in Equation~\eqref{eq:dist} stands. Essentially, we can view it as a Poisson regression model, albeit with an unusual link function.

Next, we turn to the form of the preference function. For the power preference function, and with an additional parameter for modelling purposes, we write \(g(k|\boldsymbol{\theta}_0)=k^{\alpha}+\delta\) and \(\boldsymbol{\theta}_0=(\alpha,\delta)^{\top}\), where \(\alpha,\delta>0\). The parameter \(\delta\), termed attractiveness or \emph{zero-appeal} \citep{cn06}, is so that a vertex with currently \(0\) in-degree has a positive probability of being incremented.

As mentioned in Section~\ref{intro}, under sublinear and superlinear PA, the asymptotic degree distribution is Weibull (which is considered light-tailed in EVT terminology) and essentially degenerate, respectively. On the other hand, for many real networks, the empirical degree distribution has a tail that deviates from the power law in the body but is still heavy \citep{lef24}. This means that the power preference function, while initially useful according to Figure~\ref{fig:increments-vs-degrees}, is unlikely to capture the subtle tail behaviour of real degree distributions. Therefore, we also consider the \emph{piecewise} preference function,
\begin{align}
g(k|\boldsymbol{\theta}_1)=\left\{\begin{array}{ll}
   k^{\alpha}+\delta, & k\leq \gamma, \\
   \gamma^{\alpha}+\beta(k-\gamma)+\delta, & k\geq \gamma
   \end{array}\right.,
\end{align}
where \(\boldsymbol{\theta}_1=(\alpha,\beta,\gamma,\delta)^{\top}\), and \(\beta,\gamma>0\). It increases sublinearly or superlinearly with \(k\) up to \(\gamma\) and linearly above \(\gamma\), assuming that the effect of \(\delta\) is negligible. Note that \(\gamma\) is unrelated to the power law exponent in Section~\ref{intro}. As shown by \citet{blp25}, this preference function implies a degree distribution with a heavy yet flexible tail, and is equivalently deemed \emph{tail-realistic}. Although their proof is only for when the number of new vertices \(M_t\) is fixed to \(1\), we shall use it as a practical alternative. In some applications in Section~\ref{application}, it indeed fits the data better than the power preference function does.

To derive the likelihood, we use \(\boldsymbol{\theta}_r\) to represent the general form of the parameter vector without specifying the preference function, where \(r\) can take value \(0\) or \(1\). We assume that data is collected from \(t=0\) to \(t=T\), and \(\boldsymbol{\theta}\) and \(\mu\) stay the same values throughtout, although the second assumption will be relaxed in Section~\ref{hierarchical}. Using Equation~\eqref{eq:dist}, the likelihood is
\begin{align}
  f(\mathbf{Y}|\mathbf{k},\boldsymbol{\theta}_r,\mu)&=\prod_{t=1}^T\prod_{i:i\in\mathcal{V}_{t-1}}f(Y_{i,t}|\mathbf{k},\boldsymbol{\theta}_r,\mu) \nonumber\\
  &=\prod_{t=1}^T\prod_{i:i\in\mathcal{V}_{t-1}}\frac{\exp\left[-\mu\tilde{g}(k_{i,t-1}|\boldsymbol{\theta}_r)\right]\left[\mu\tilde{g}(k_{i,t-1}|\boldsymbol{\theta}_r)\right]^{Y_{i,t}}}{Y_{i,t}!} \nonumber\\
  &\propto \mu^{A_T(\mathbf{Y})}\exp\left[-\mu{}\sum_{t=1}^T\underbrace{\sum_{i:i\in\mathcal{V}_{t-1}}\tilde{g}(k_{i,t-1}|\boldsymbol{\theta}_r)}_{\text{sum to 1}}\right]B_T(\mathbf{k},\mathbf{Y},\boldsymbol{\theta}_r),\nonumber\\
  &= \mu^{A_T(\mathbf{Y})}\exp\left(-\mu{}T\right)B_T(\mathbf{k},\mathbf{Y},\boldsymbol{\theta}_r),\nonumber
\end{align}
where \(A_T(\mathbf{Y}):=\sum_{t=1}^T\sum_{i:i\in\mathcal{V}_{t-1}}Y_{i,t}\), and \(B_T(\mathbf{k},\mathbf{Y},\boldsymbol{\theta}_r):= \prod_{t=1}^T\prod_{i:i\in\mathcal{V}_{t-1}}\tilde{g}(k_{i,t-1}|\boldsymbol{\theta}_r)^{Y_{i,t}}\).

While the parameter inference using the Bayesian approach is deferred to Section~\ref{inference}, it is useful to assign to \(\mu\) a Gamma\((a,b)\) prior here, where \(b\) is the rate parameter, so that \(\mu\) can be integrated out analytically:
\begin{align}
f(\mathbf{Y}|\mathbf{k},\boldsymbol{\theta}_r)
  &=\int_0^{\infty}\pi(\mu)\times{}f(\mathbf{Y}|\mathbf{k},\boldsymbol{\theta}_r,\mu)d\mu\nonumber\\
  &\propto\int_0^{\infty}\frac{b^a}{\Gamma(a)}\mu^{a-1}\exp\left(-b\mu\right)\times\mu^{A_T(\mathbf{Y})}\exp\left(-\mu{}T\right){}B_T(\mathbf{k},\mathbf{Y},\boldsymbol{\theta}_r)d\mu\nonumber\\
  &=B_T(\mathbf{k},\mathbf{Y},\boldsymbol{\theta}_r)\frac{b^a}{\Gamma(a)}{}\int_0^{\infty}\mu^{a+A_T(\mathbf{Y})-1}\exp\left[-\mu\left(b+T\right)\right]d\mu\nonumber\\
  &=B_T(\mathbf{k},\mathbf{Y},\boldsymbol{\theta}_r)\frac{b^a \Gamma(a+A_T(\mathbf{Y}))}{\Gamma(a) (b+T)^{-(a+A_T(\mathbf{Y}))}}.
  \label{eq:integrate}
\end{align}
The fraction does not involve \(\boldsymbol{\theta}_r\) but will become useful when the hyperparameters \(a\) and \(b\) are, with a reprarametrisation, assigned with hyperpriors and to be inferred in Section~\ref{hierarchical}.

\hypertarget{inference}{%
\section{Bayesian inference}\label{inference}}

In this section, we outline the Bayesian inference procedure, in which the selection between the preference functions is incorporated.

Using Equation~\eqref{eq:integrate} and by assigning relatively uninformative and independent priors to the components of \(\boldsymbol{\theta}_r\), denoted collectively by \(\pi(\boldsymbol{\theta}_r)\), we obtain its posterior up to a proportionality constant:
\begin{align}
\pi(\boldsymbol{\theta}_r|\mathbf{k},\mathbf{Y})&\propto \pi(\boldsymbol{\theta}_r)f(\mathbf{Y}|\mathbf{k},\boldsymbol{\theta}_r)=\pi(\boldsymbol{\theta}_r)B_T(\mathbf{k},\mathbf{Y},\boldsymbol{\theta}_r)\frac{b^a \Gamma(a+A_T(\mathbf{Y}))}{\Gamma(a) (b+T)^{-(a+A_T(\mathbf{Y}))}}.
  \label{eq:posterior}
\end{align}
Inference is carried out by drawing samples from \(\pi(\boldsymbol{\theta}_r|\mathbf{k},\mathbf{Y})\), using Markov chain Monte Carlo (MCMC). As \(\boldsymbol{\theta}_r\) has a very low dimension of \(2r+2\), a simple sampler such as Metropolis-Hastings suffices. We term the model associated with Equation~\eqref{eq:posterior} the \emph{single} model, as opposed to the \emph{hierarchical} model introduced in Section~\ref{hierarchical}.

\hypertarget{model-select}{%
\subsection{Selecting between the preference functions}\label{model-select}}

While PA in general can be evidenced from the diagnostic proposed in Section~\ref{preliminaries} and illustrated in Section~\ref{eda}, for a particular set of data, it might not be clear whether the power preference function \(g(\cdot|\boldsymbol{\theta}_0)\) or the piecewise preference function \(g(\cdot|\boldsymbol{\theta}_1)\) is more appropriate, even though the latter is argued to be tail-realistic. Such issue can however be resolved by framing this as a \emph{model selection} problem, and calculating the evidence of one preference function over the other. As far as the comparison between the preference functions is concerned, we drop the word ``model'' to avoid confusion. This selection procedure can be embedded as an additional step in the MCMC sampler. Among the various methods to embed the selection step, we use that by \citet{cb95}, which is outlined below.

First, we make \(r\) a Bernoulli random variable with prior \(\pi(r)=p^r{}(1-p)^{1-r}\), where \(p\) is the prior probability of piecewise preference function. Next, we compute the joint posterior of \(\boldsymbol{\theta}_r\) and \(r\) up to a proportionality constant, i.e.~\(\pi(\boldsymbol{\theta}_r,r|\mathbf{k},\mathbf{Y})\propto\pi(r)\pi(\boldsymbol{\theta}_r)f(\mathbf{Y}|\mathbf{k},\boldsymbol{\theta}_r)\). Also, we specify a \emph{pseudoprior} of \((\beta,\gamma)^{\top}\) under \(r=0\), denoted by \(\pi^{*}(\beta,\gamma)\), with the prefix ``pseudo'' due to that \(\beta\) and \(\gamma\) do not exist in the parametrisation of the power preference function. The choice of \(\pi^{*}(\beta,\gamma)\) theoretically does not affect \(\pi(r|\mathbf{k},\mathbf{Y})\) but only the efficiency of the MCMC sampler. Now, we sample from the joint posterior by repeating the following steps:

\begin{enumerate}
\def\labelenumi{\arabic{enumi}.}
\tightlist
\item
  Draw from Equation~\eqref{eq:posterior} a sample of \(\boldsymbol{\theta}_r\), denoted by \((\alpha^{*},\delta^{*})^{\top}\) if the current value of \(r\) is \(0\), and \((\alpha^{*},\beta^{*},\gamma^{*},\delta^{*})^{\top}\) if the current value of \(r\) is \(1\).
\item
  If the current value of \(r\) is \(0\), also draw a sample from \(\pi^{*}(\beta,\gamma)\), denoted by \((\beta^{*},\gamma^{*})^{\top}\).
\item
  Regardless of the value of \(r\), write \(\boldsymbol{\theta}^{*}_0=(\alpha^{*},\delta^{*})^{\top}\) and \(\boldsymbol{\theta}^{*}_1=(\alpha^{*},\beta^{*},\gamma^{*},\delta^{*})^{\top}\).
\item
  Compute \(P_0=\pi(\boldsymbol{\theta}^{*}_0|\mathbf{k},\mathbf{Y})\pi(\boldsymbol{\theta}^{*}_0)\pi^{*}(\beta^{*},\gamma^{*})(1-p)\) and \(P_1=\pi(\boldsymbol{\theta}^{*}_1|\mathbf{k},\mathbf{Y})\pi(\boldsymbol{\theta}^{*}_1)p\).
\item
  With probability \(\frac{P_0}{P_0+P_1}\), set \(r=0\) and \(\boldsymbol{\theta}_r=\boldsymbol{\theta}_0^{*}\); otherwise, set \(r=1\) and \(\boldsymbol{\theta}_r=\boldsymbol{\theta}_1^{*}\).
\end{enumerate}

The proportions of \(r=0\) and \(r=1\) in the MCMC samples will be estimates of the posterior probabilities \(\pi(r=0|\mathbf{k},\mathbf{Y})\) and \(\pi(r=1|\mathbf{k},\mathbf{Y})\), respectively. They are then used to obtain the estimate of the Bayes factor \(B_{10}:=\frac{\pi(r=1|\mathbf{k},\mathbf{Y})(1-p)}{\pi(r=0|\mathbf{k},\mathbf{Y})p}\), which tells us the evidence of piecewise preference function over the power preference function.

One should note that \(p\) and \(1-p\) are often specified not to represent the prior belief on the preference function, but to practically facilitate the selection. This is because assigning equal probabilities \emph{a priori} to the two preference functions could result in one of them not being visited very often when running the MCMC sampler, and subsequently poor estimation of the Bayes factor, as in the case of e.g.~\citet{fp08}. Therefore, when necessary, unequal prior probabilities are assigned in the applications in Section~\ref{application}.

Lastly, and relatedly, in some instances in Section~\ref{application}, \(\delta\) will be dropped or equivalently fixed to \(0\). The inference procedures still stand, with \(\boldsymbol{\theta}_0\) and \(\boldsymbol{\theta}_1\) becoming \(\alpha\) and \((\alpha,\beta,\gamma)^{\top}\), respectively. While it is possible to incorporate the inclusion or exclusion of \(\delta\) in the selection so that \(r\) can take \(4\) values corresponding to \(\boldsymbol{\theta}_r=\alpha\), \((\alpha,\delta)^{\top}\), \((\alpha,\beta,\gamma)^{\top}\), \((\alpha,\beta,\gamma,\delta)^{\top}\), we deem this not necessary.

\hypertarget{hierarchical}{%
\section{Hierarchical modelling}\label{hierarchical}}

When the observation period (\(T\)) is sufficiently large, it may not be appropriate to assume that \(\boldsymbol{\theta}_r\) stays the same throughout. It is therefore natural to split into shorter periods, assume that the parameters are constant within each, and allow variability across periods. This can be facilitated by turning the single model into its hierarchical version, which is the goal of this section.

We first update our notation by assuming that there are \(S\) periods, and \(T_s\) time points in period \(s=1,2,\ldots,S\). As in our CRAN dependencies data, a usually natural choice of a period is a month. For notational convenience, \(\mathbf{k}_s\) represents the in-degrees within period \(s\) such at \(\mathbf{k}=(\mathbf{k}_1,\mathbf{k}_2,\ldots,\mathbf{k}_S)\), and such convention applies to \(\mathbf{X}\), \(\mathbf{Y}\) and \(\mathbf{Z}\). Next, we reparametrise the hyperparameters for \(\mu\) by replacing \(a\) and \(b\) by \(\mu_{\mu}^2/\sigma_{\mu}^2\) and \(\mu_{\mu}/\sigma_{\mu}^2\), respectively, so that the expectation and variance are \(\mu_{\mu}\) and \(\sigma_{\mu}^2\), respectively. Now, as the single model is assumed to apply to the increments within each period \(s\), Equation~\eqref{eq:integrate} is rewritten according to the reparametrisation and subscripting the remaining quantities:
\begin{align}
f(\mathbf{Y}_s|\mathbf{k},\boldsymbol{\theta}_{rs},\mu_{\mu},\sigma_{\mu})&\propto{}B_{T_s}(\mathbf{k}_s,\mathbf{Y}_s,\boldsymbol{\theta}_{rs})\frac{\left(\mu_{\mu}/\sigma_{\mu}^2\right)^{\mu_{\mu}^2/\sigma_{\mu}^2}\Gamma\left(\mu_{\mu}^2/\sigma_{\mu}^2+A_{T_s}(\mathbf{Y}_s)\right)}{\Gamma\left(\mu_{\mu}^2/\sigma_{\mu}^2\right)\left(\mu_{\mu}/\sigma_{\mu}^2+T_s\right)^{-\left(\mu_{\mu}^2/\sigma_{\mu}^2+A_{T_s}(\mathbf{Y}_s)\right)}}.
\label{eq:likelihood-hier}
\end{align}
We assign independent normal priors to each component of \(\boldsymbol{\theta}_{rs}\), for \(s=1,2,\ldots,S\). Collectively, we write \(\boldsymbol{\theta}_{rs}\overset{\text{iid}}{\sim}\pi(\cdot|\boldsymbol{\phi}_r)\), where \(\boldsymbol{\phi}_r\) is the vector of hyperparameters. To complete the model specification, we assign independent normal and half-Cauchy hyperpriors to the mean and standard deviation hyperparameters, respectively, with the hyperprior denoted by \(\pi(\boldsymbol{\phi}_r,\mu_{\mu},\sigma_{\mu})\) collectively.

Using \(r=1\) for illustration for the rest of this section, as the components of the parameter vector \(\boldsymbol{\theta}_{rs}=(\alpha_s,\beta_s,\gamma_s,\delta_s)^{\top}\) are independently and normally distributed \emph{a priori},
\begin{align}
\boldsymbol{\theta}_{rs}&\overset{\text{iid}}{\sim}\text{MVN}\left((\mu_{\alpha},\mu_{\beta},\mu_{\gamma},\mu_{\delta})^{\top},\text{diag}(\sigma_{\alpha}^2,\sigma_{\beta}^2,\sigma_{\gamma}^2,\sigma_{\delta}^2)\right),\qquad{}s=1,2,\ldots,S,
\label{eq:prior}
\end{align}
where ``MVN'' and ``diag'' stand for the multivariate normal distribution and the diagonal matrix, respectively. Regarding the hyperparameters \(\boldsymbol{\phi}_r=\left(\mu_{\alpha},\sigma_{\alpha},\mu_{\beta},\sigma_{\beta},\mu_{\gamma},\sigma_{\gamma},\mu_{\delta},\sigma_{\delta}\right)^{\top}\), \(\mu_{\mu}\) and \(\sigma_{\mu}\),
\begin{align}
\mu_{\eta}&\sim\text{N}(m_{\eta},s_{\eta}^2),~\pi(\sigma_{\eta})=\frac{2}{\pi{}r_{\eta}\left[1+(\sigma_{\eta}/r_{\eta})^2\right]}\mathbf{1}_{\{\sigma_{\eta}>0\}},\qquad{}\eta=\alpha,\beta,\gamma,\delta,\mu,
\label{eq:hyperprior}
\end{align}
where \(\mathbf{1}_{\{E\}}\) is \(1\) if event \(E\) happens, \(0\) otherwise, and \(m_{\eta}\), \(s_{\eta}\) and \(r_{\eta}\) are pre-specified. Note that \(s_{\eta}\) is unrelated to \(s\), the index for the period, and \(r_{\eta}\) is unrelated to \(r\), the preference function variable. The joint posterior of all parameters \(\boldsymbol{\Theta}_r:=(\boldsymbol{\theta}_{r1},\boldsymbol{\theta}_{r2},\ldots,\boldsymbol{\theta}_{rS})\) and the hyperparameters, up to a proportionality constant, is therefore
\begin{align}
\pi(\boldsymbol{\Theta}_r,\boldsymbol{\phi}_r,\mu_{\mu},\sigma_{\mu}|\mathbf{k},\mathbf{Y})&\propto\pi(\mathbf{Y},\boldsymbol{\Theta}_r,\boldsymbol{\phi}_r,\mu_{\mu},\sigma_{\mu}|\mathbf{k})\nonumber\\
&=f(\mathbf{Y}|\mathbf{k},\boldsymbol{\Theta}_r,\mu_{\mu},\sigma_{\mu})\pi(\boldsymbol{\Theta}_r|\boldsymbol{\phi}_r)\times\pi(\boldsymbol{\phi}_r,\mu_{\mu},\sigma_{\mu})\nonumber\\
&=\left[\prod_{s=1}^Sf(\mathbf{Y}_s|\mathbf{k}_s,\boldsymbol{\theta}_{rs},\mu_{\mu},\sigma_{\mu})\pi(\boldsymbol{\theta}_{rs}|\boldsymbol{\phi}_r)\right]\times\pi(\boldsymbol{\phi}_r,\mu_{\mu},\sigma_{\mu}),
\label{eq:posterior-hier}
\end{align}
where the first, second and last terms are according to Equations~\eqref{eq:likelihood-hier}, \eqref{eq:prior} and \eqref{eq:hyperprior}, respectively. While the dimension is \(4S+10\) when \(r=1\) and \(2S+6\) when \(r=0\), a Metropolis-Hastings sampler still suffices when \(r\) is fixed, as the posterior is fully tractable. On the other hand, as selecting between the preference functions amounts to a difference of \(2S+4\) in the number of parameters, the steps outlined in Section~\ref{model-select} are unlikely to be useful even when \(S\) is moderately large. Therefore, no selection is considered when applying the hierarchical model according to Equation~\eqref{eq:posterior-hier}.

\hypertarget{application}{%
\section{Application}\label{application}}

In this section, we show the results of fitting the single and hierarchical models to the CRAN dependencies data explored in Section~\ref{eda}. As the results of both models applied to both preference functions are compared, for convenience we will use the parameters without any subscripts. Each combination of the model and the preference function was applied to the three categories of increments (internal, external, deletion) and the two types of dependencies (\texttt{Imports} and \texttt{Imports} \(+\) \texttt{Depends}). For the single model, after discarding the first 1000 iterations and recording the values every 10 iterations, an MCMC sample of size 10000 was obtained. For the hierarchical model, after discaring the first 10000 iterations and recording the values every 20 iterations, an MCMC sample of size 10000 was obtained. Convergence was checked and the effective sample sizes, which are reported in the appendices, were deemed sufficient.

\begin{figure}[!pt]

{\centering \includegraphics[width=0.95\linewidth]{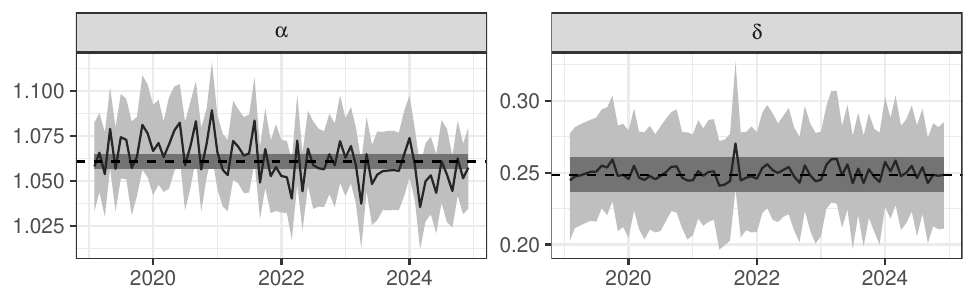} 

}

\caption{Posterior mean (solid) and 95\% credible intervals (hovering band) of the power preference function parameters under the hierarchical model, fitted to the external increments of \texttt{Imports} between CRAN packages. Overlaid are the posterior mean (dashed) and 95\% credible intervals (horizontal band) under the single model.}\label{fig:imports-external-power}
\end{figure}

We first look at the results for one category-type combination. In Figure~\ref{fig:imports-external-power}, we plot the posterior of the power preference function parameters, under both the single and hierarchical models, fitted to the external increments of \texttt{Imports}. Alignment is observed as, while the hierarchical model indicates fluctuations over time, its posterior mean hovers around the single model counterpart, albeit with more uncertainty. For \(\alpha\), the estimates are consistently above \(1\), suggesting superlinear PA. For \(\delta\), there is a blip around September 2021, indicating a larger zero-appeal for packages at that time.

\begin{figure}[!pb]

{\centering \includegraphics[width=0.95\linewidth]{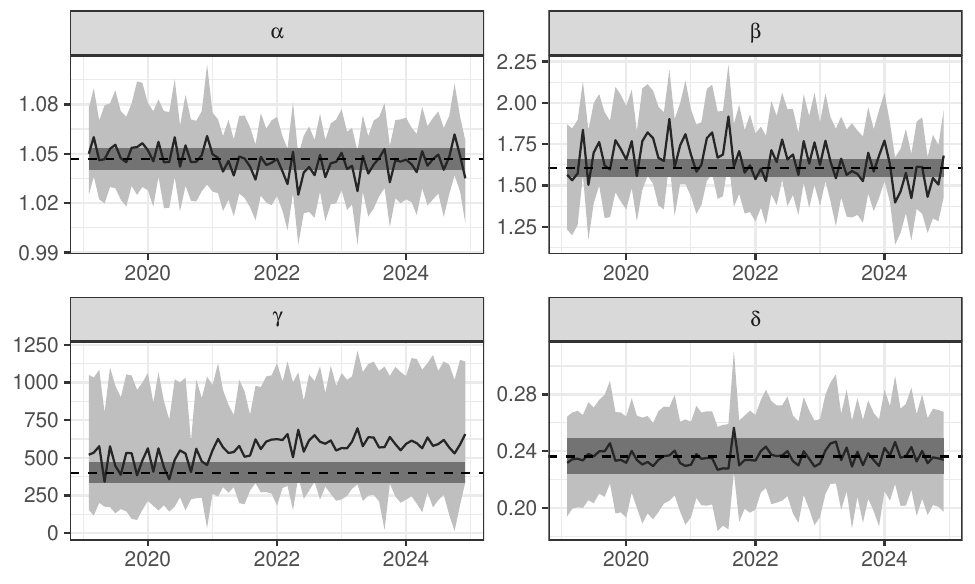} 

}

\caption{Posterior mean (solid) and 95\% credible intervals (hovering band) of the piecewise preference function parameters under the hierarchical model, fitted to the external increments of \texttt{Imports} between CRAN packages. Overlaid are the posterior mean (dashed) and 95\% credible intervals (horizontal band) under the single model.}\label{fig:imports-external-piecewise}
\end{figure}

We plot the counterparts for the piecewise preference function in Figure~\ref{fig:imports-external-piecewise}, with several interesting observations. First, alignment between the single and hierarchical models is largely retained. Second, with slightly larger uncertainty, the estimates for \(\alpha\) are still consistently above \(1\), barring a couple of months. Third, uncertainty is quite substantial for \(\gamma\), which is the threshold of the piecewise preference function, under the hierarchical model, considering the range of in-degrees in Figures~\ref{fig:increments-vs-degrees} and \ref{fig:example-degrees}. Fourth, comparing Figures~\ref{fig:imports-external-power} and \ref{fig:imports-external-piecewise}, the pattern of the estimates for \(\delta\) is highly similar under the two preference functions, suggesting that the blip around September 2021 is likely a genuine shift in how dependencies behaved at that time.

\begin{figure}[!hp]

{\centering \includegraphics[width=0.95\linewidth]{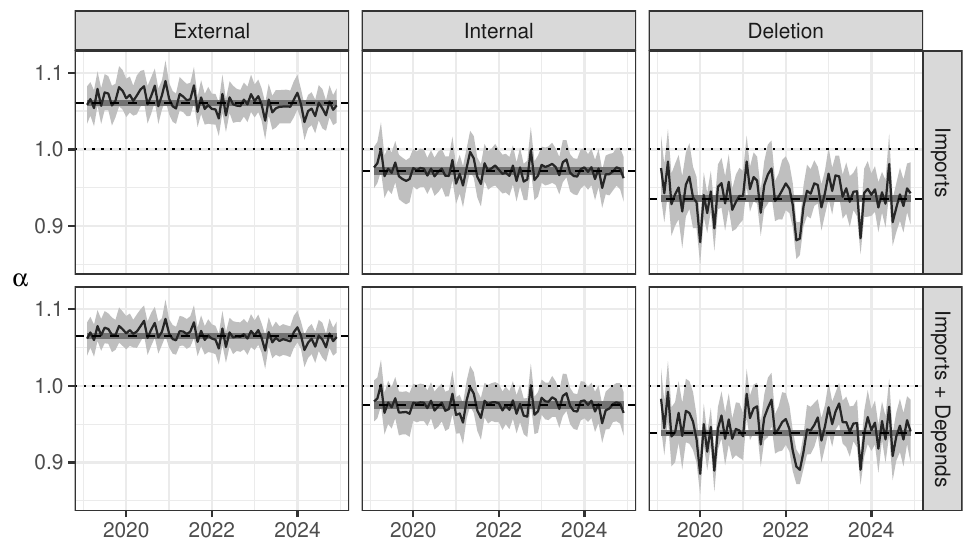} \includegraphics[width=0.95\linewidth]{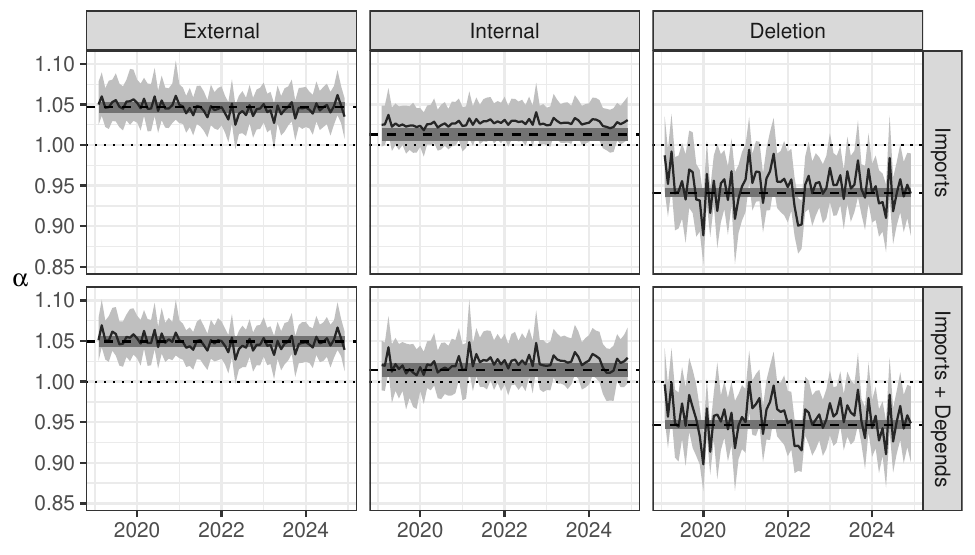} 

}

\caption{Posterior mean (solid) and 95\% credible intervals (hovering band) of $\alpha$ of the power (top 2 rows) and piecewise (bottom 2 rows) preference function under the hierarchical model, fitted to all category-type combinations of dependencies between CRAN packages. Overlaid are the posterior mean (dashed) and 95\% credible intervals (horizontal band) under the single model. The dotted line $\alpha=1$ represents linear PA.}\label{fig:alpha}
\end{figure}

Looking at the broader set of results, we plot the estimates of \(\alpha\) across all category-type combinations, by both single and hierarchical models, and for both preference functions, in Figure~\ref{fig:alpha}. Holding other things constant, the difference between the two types of dependencies is negligible. As this similarity applies to other parameters, the type \texttt{Imports} \(+\) \texttt{Depends} is not included in the remaining plots of this section but instead shown in the appendices. On the other hand, there are noticeable differences between the categories, with some interaction with the preference function. Specifically, external and deletion increments suggest superlinear and sublinear PA, respectively, under either the power or piecewise preference function, albeit partially for the latter. On the other hand, the estimates of \(\alpha\) go from mostly below \(1\) for the power preference function, to mostly above \(1\) for the piecewise preference function, suggesting a difference in the fit that is subtle yet significant for the implications on the tail of the degree distribution.

\begin{figure}[!ht]

{\centering \includegraphics[width=0.75\linewidth]{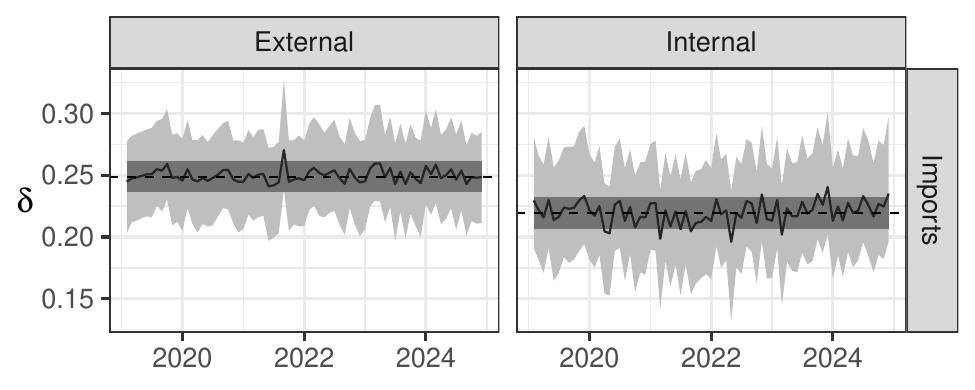} \includegraphics[width=0.75\linewidth]{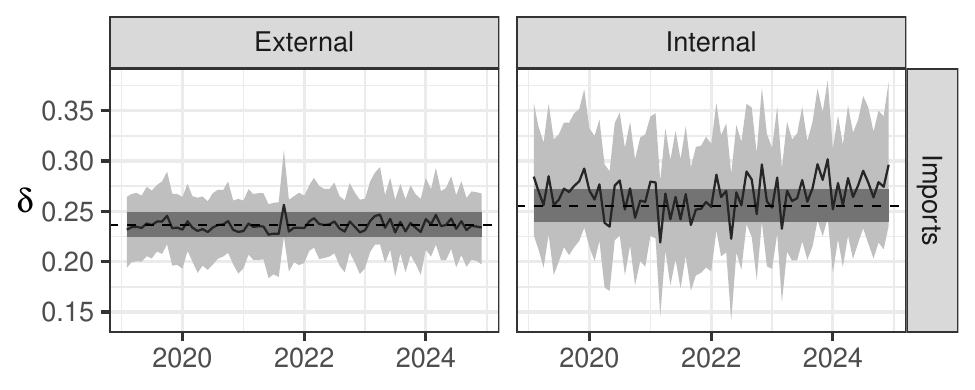} 

}

\caption{Posterior mean (solid) and 95\% credible intervals (hovering band) of $\delta$ of the power (top) and piecewise (bottom) preference function under the hierarchical model, fitted to the external and internal increments of \texttt{Imports} between CRAN packages. Overlaid are the posterior mean (dashed) and 95\% credible intervals (horizontal band) under the single model.}\label{fig:delta}
\end{figure}

Next, we look at the results for \(\delta\) in Figure~\ref{fig:delta}. When fitting to the deletion increments, the MCMC performs poorly, in the sense that it takes many more iterations to obtain a representative number of samples. Closer examination reveals that many samples of \(\delta\) are very close to \(0\) and with a magnitude much lower than those in Figures~\ref{fig:imports-external-power} and \ref{fig:imports-external-piecewise}. As this aligns with the interpretation that a package does not need a positive zero-appeal to have a dependency (on that package) deleted, \(\delta\) is therefore removed from the model, or equivalent fixed to \(0\), in the fits to the deletion increments, hence the omission in the Figure~\ref{fig:delta}. Looking at the remaining two categories, the behaviour stays largely the same across the types and preference functions for external increments. On the other hand, for the internal increments, \(\delta\) is seemingly larger under the piecewise preference function than under the power preference function.

\begin{figure}[!ht]

{\centering \includegraphics[width=0.95\linewidth]{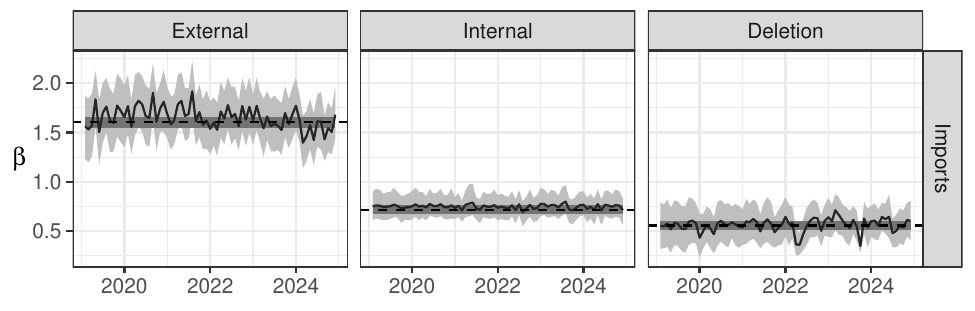} \includegraphics[width=0.95\linewidth]{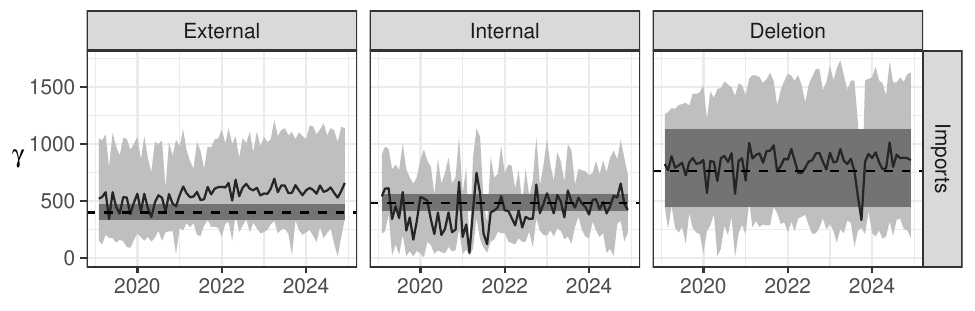} 

}

\caption{Posterior mean (solid) and 95\% credible intervals (hovering band) of $\beta$ (top) and $\gamma$ (bottom) of the piecewise preference function under the hierarchical model, fitted to all 3 categories of increments of \texttt{Imports} between CRAN packages. Overlaid are the posterior mean (dashed) and 95\% credible intervals (horizontal band) under the single model.}\label{fig:beta-gamma}
\end{figure}

For the two parameters \(\beta\) and \(\gamma\) that only exist in the piecewise preference function, we plot their estimates in Figure~\ref{fig:beta-gamma}, which again suggests alignment between single and hierarchical models, yet differing behaviours across categories. The substantial uncertainty for \(\gamma\) of both single and hierarchical models applied to the deletion increments indicates difficulty in locating the threshold of the piecewise preference function. This will be revisited when determining which preference function fits better for each set of data.

\begin{table}[H]

\caption{\label{tab:model-selection}Bayes factor of piecewise preference function over power preference function for all category-type combinations of dependencies between CRAN packages. A number greater (smaller) than $1$ means that the former (latter) is preferred, and how far away it is from $1$ indicates the strength of such evidence.}
\centering
\begin{tabular}[t]{l|r|r|r}
\hline
  & External & Internal & Deletion\\
\hline
Imports & 2003.44 & > 9.999e+13 & 0.0008617\\
\hline
Imports + Depends & 1021240.00 & > 9.999e+13 & 0.2795490\\
\hline
\end{tabular}
\end{table}

\hypertarget{selecting-between-the-preference-functions}{%
\subsection{Selecting between the preference functions}\label{selecting-between-the-preference-functions}}

Using the procedure in Section~\ref{model-select}, we calculate the Bayes factors of the piecewise preference function over the power preference function, which are reported Table~\ref{tab:model-selection}. For external increments, it is clear that the former is preferred. For internal increments, with \(p\) (prior probability of the former) being as small as \(10^{-10}\), all 10000 MCMC samples of \(r\) are of value \(1\), meaning that the estimated Bayes factor is greater than \(\frac{(10000 - 1) \times(1-10^{-10})}{1 \times 10^{-10}}\approx \ensuremath{9.999\times 10^{13}}\), and that the evidence for the former is even stronger. On the other hand, for deletion increments, the evidence for the power preference function is moderate to strong, which complements the huge uncertainty for \(\gamma\) under the piecewise preference function shown in Figure~\ref{fig:beta-gamma}. Coupled with that \(\delta\) is excluded in the fits, it is suggested that, for the deletion increments, a single parameter \(\alpha\) suffices for describing the sublinear PA (or, more precisely, preferential \emph{detachment}) behaviour.

\begin{figure}[!hp]

{\centering \includegraphics[width=0.95\linewidth]{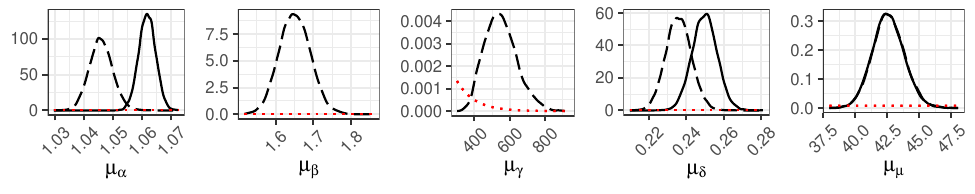} \includegraphics[width=0.95\linewidth]{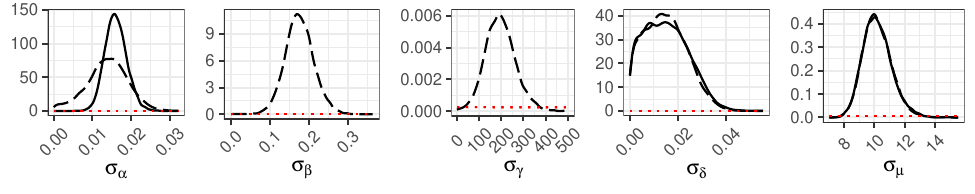} \includegraphics[width=0.95\linewidth]{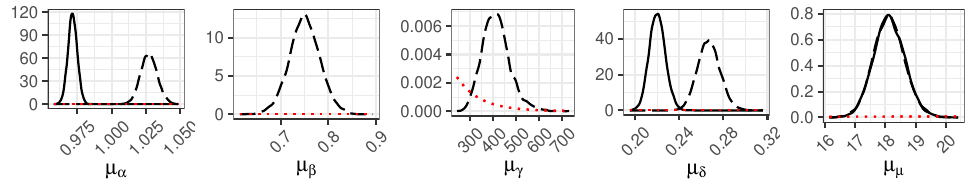} \includegraphics[width=0.95\linewidth]{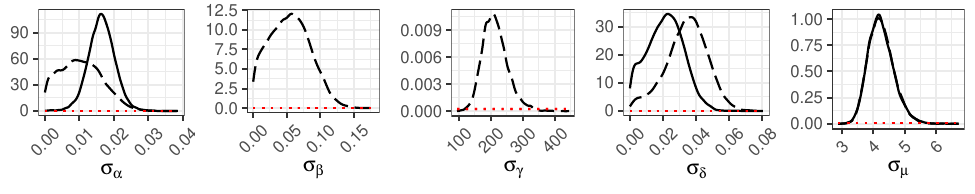} \includegraphics[width=0.95\linewidth]{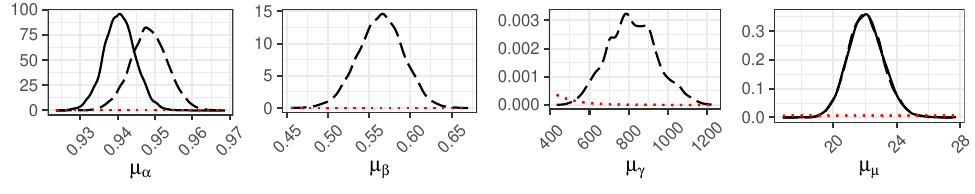} \includegraphics[width=0.95\linewidth]{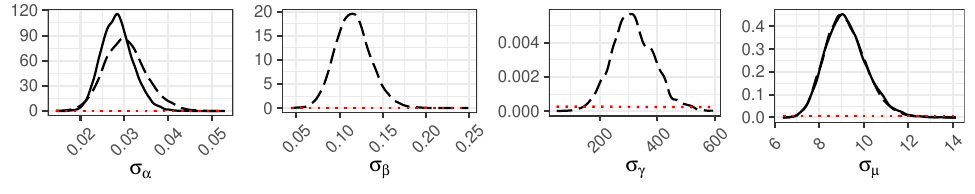} 

}

\caption{Posterior density of the hierarchical model hyperparameters under the power (solid) and piecewise (dashed) preference function, applied to the external (top 2 rows), internal (middle 2 rows) and deletion (bottom 2 rows) increments of \texttt{Imports} between CRAN packages. The dotted line is the prior density.}\label{fig:prior-posterior-08}
\end{figure}

\hypertarget{posterior-of-the-hierarchical-model-hyperparameters}{%
\subsection{Posterior of the hierarchical model hyperparameters}\label{posterior-of-the-hierarchical-model-hyperparameters}}

The posterior densities of the hierarchical model hyperparameters, for both preference functions and all three categories of increments of \texttt{Imports}, are plotted in Figure~\ref{fig:prior-posterior-08}. The results for \texttt{Imports} \(+\) \texttt{Depends} are similar and therefore not shown. The flatness of the dotted line overlaid suggests that the prior has very little influence on the posterior of all the hyperparameters (except for perhaps \(\mu_{\gamma}\)), mainly because of the informativeness of the data. A noticeable difference between the two preference functions is in their posterior densities for \(\mu_{\alpha}\), which are not only non-overlapping but also on opposite sides of \(1\) for the fit to internal increments (third row, leftmost plot). This is consistent with the observations on \(\alpha\) in the middle column of Figure~\ref{fig:alpha}.

\hypertarget{discussion}{%
\section{Discussion}\label{discussion}}

In this paper, we propose a model for regressing the increments on the in-degrees, with the aim to accurately measure preferential attachment (PA) as the network evolves and grows. The results of fitting the single and hierarchical versions of the model to the CRAN dependencies data show broad alignment, with the parameters under the latter being relatively stable over the observation period. Consideration has also been given to the preference function, with the selection between the commonly used power preference function and the tail-realistic piecewise preference function embedded in the MCMC sampler. Such selection reveals subtly different behaviours between the addition and deletion of edges. For the former, no matter internal or external, there is strong evidence for the piecewise preference function with \(\alpha>1\), implying superlinear PA up to a threshold (\(\gamma\)) and linear PA beyond \(\gamma\). For the latter, there is moderate to strong evidence for the power preference function with \(\alpha<1\) and \(\delta\) fixed to \(0\), implying sublinear PA at any in-degree and without zero-appeal.

The usefulness of the piecewise preference function addresses the paradox of the co-existence of superlinear preference and scale-freeness \citep{kk08}. The plausible explanation is that real networks with such properties grow superlinearly but only up to a certain point (\(\gamma\)), leading to partial scale-freeness without contradiction.

As a natural extension of the model, joint modelling can be considered. For example, the increments can be modelled jointly with the degrees, as the theoretical degree distribution can be expressed in terms of the preference function parameters, using the results by \citet{rtv07}. This however will require the assumption that the network has been growing according to the same set of parameter values, meaning that the single model would be more applicable. The three categories could also be modelled jointly, although different rates of PA revealed in this paper have to be taken into account.

When applying to a broader set of real networks, other network properties can be jointly modelled with PA. First, reciprocity, which is by nature absent in our data set but empirically observed in many others, can be incorporated by modelling the in-degrees and out-degrees simultaneously using the approach by \citet{cw24} and \citet{wr24}. Second, the fitness of the vertex could be incorporated in a way similar to \citet{pss16} and \citet{obu19}, and untangled from the PA induced by the in-degree. The rich information in the increments data increases the possibility of inferring the fitness alongside other parameters. Third, the aging effect, which has been incorporated by \citet{zzlw08} and \citet{lzcxa13} in their analyses of PA in software dependencies, could also be considered. Lastly, while clustering attachment \citep{mrv26} could be considered, that it implies a light-tailed degree distribution warrants modifications in the model.

More flexible preference functions can be considered, such as one where the smallest degrees (\(1\), \(2\) and perhaps \(3\)) do not follow the general form. The mixture proposed by \citet{amc21} and \citet{arnold21}, can also be used to discriminate between PA and uniform attachment. However, one has to look out for potential identifiability issues if the piecewise preference function, which already brings about flexibility, is used as a mixture component. Orthogonally, for the hierarchical model, rather than being constant for each period, each parameter of the preference function, regardless of the specific form, can be assumed to change smoothly over time and potentially arise from a Gaussian process prior.

\hypertarget{funding-acknowledgements}{%
\section*{Funding Acknowledgements}\label{funding-acknowledgements}}
\addcontentsline{toc}{section}{Funding Acknowledgements}

This research received no specific grant from any funding agency in the public, commerical, or not-for-profit sectors.

\hypertarget{disclosure-statement}{%
\section*{Disclosure Statement}\label{disclosure-statement}}
\addcontentsline{toc}{section}{Disclosure Statement}

The authors report there are no competing interests to declare.

\hypertarget{data-availability-statement}{%
\section*{Data Availability Statement}\label{data-availability-statement}}
\addcontentsline{toc}{section}{Data Availability Statement}

The authors confirm that the data supporting the findings of this study are available within the article or its supplementary materials.

\bibliographystyle{apalike}
\renewcommand\refname{References}
\bibliography{ref-mix.bib}

\begin{thebibliography}{}

\bibitem[Arnold, 2021]{arnold21}
Arnold, N. (2021).
\newblock {\em Studying evolving complex networks}.
\newblock PhD thesis, Queen Mary University of London.

\bibitem[Arnold et~al., 2021]{amc21}
Arnold, N.~A., Mondrag\'{o}n, R.~J., and Clegg, R.~G. (2021).
\newblock Likelihood-based approach to discriminate mixtures of network models
  that vary in time.
\newblock {\em Scientific Reports}, 11.

\bibitem[Artico et~al., 2020]{asvw20}
Artico, I., Smolyarenko, I., Vinciotti, V., and Wit, E.~C. (2020).
\newblock How rare are power-law networks really?
\newblock {\em Proceedings of the Royal Society A: Mathematical, Physical and
  Engineering Sciences}, 476(2241):20190742.

\bibitem[Barab\'{a}si and Albert, 1999]{ba99a}
Barab\'{a}si, A.-L. and Albert, R. (1999).
\newblock Emergence of scaling in random networks.
\newblock {\em Science}, 286(5439):509--512.

\bibitem[Bollob\'{a}s et~al., 2003]{bbcr03}
Bollob\'{a}s, B., Borgs, C., Chayes, J., and Riordan, O. (2003).
\newblock Directed scale-free graphs.
\newblock In {\em Proceedings of the Fourteenth Annual ACM-SIAM Symposium on
  Discrete Algorithms}, pages 132--139, New York. ACM.

\bibitem[Bollob\'{a}s et~al., 2001]{brst01}
Bollob\'{a}s, B., Riordan, O., Spencer, J., and Tusn\'{a}dy, G. (2001).
\newblock The degree sequence of a scale-free random graph process.
\newblock {\em Random Structures Algorithms}, 18(3):279--290.

\bibitem[Boughen et~al., 2025]{blp25}
Boughen, T., Lee, C., and {Palacios {R}amirez}, V. (2025).
\newblock Tail flexibility in the degrees of preferential attachment networks.
\newblock {\em ArXiv e-prints}.
\newblock arXiv:2506.18726.

\bibitem[Broido and Clauset, 2019]{bc19}
Broido, A.~D. and Clauset, A. (2019).
\newblock Scale-free networks are rare.
\newblock {\em Nature Communications}, 10(1017).

\bibitem[Carlin and Chib, 1995]{cb95}
Carlin, B.~P. and Chib, S. (1995).
\newblock Bayesian model choice via {M}arkov chain {M}onte {C}arlo methods.
\newblock {\em Journal of the Royal Statistical Society: Series B
  (Methodological)}, 57(3):473--484.

\bibitem[Charpentier and Flachaire, 2019]{cf19}
Charpentier, A. and Flachaire, E. (2019).
\newblock Extended scale-free networks.
\newblock {\em ArXiv e-prints}.
\newblock arXiv:1905.10267.

\bibitem[Cirkovic and Wang, 2024]{cw24}
Cirkovic, D. and Wang, T. (2024).
\newblock Modeling random networks with heterogeneous reciprocity.
\newblock {\em Journal of Machine Learning Research}, 25:1--40.

\bibitem[Csardi and Nepusz, 2006]{cn06}
Csardi, G. and Nepusz, T. (2006).
\newblock The igraph software package for complex network research.
\newblock {\em Inter{J}ournal}, Complex Systems:1695.

\bibitem[Dorogovtsev and Mendes, 2002]{dm02}
Dorogovtsev, S.~N. and Mendes, J. F.~F. (2002).
\newblock Evolution of networks.
\newblock {\em Advances in Physics}, 51(4):1079--1187.

\bibitem[Friel and Pettitt, 2008]{fp08}
Friel, N. and Pettitt, A.~N. (2008).
\newblock Marginal likelihood estimation via power posteriors.
\newblock {\em Journal of the Royal Statistical Society: Series B (Statistical
  Methodology)}, 70(3):589--607.

\bibitem[Gupta and Porter, 2022]{gp22}
Gupta, H. and Porter, M.~A. (2022).
\newblock {Mixed logit models and network formation}.
\newblock {\em Journal of Complex Networks}, 10(6):cnac045.

\bibitem[Hasheminezhad et~al., 2020]{hbb20}
Hasheminezhad, R., Boudourides, M., and Brandes, U. (2020).
\newblock Scale-free networks need not be fragile.
\newblock In {\em 2020 {IEEE}/{ACM} International Conference on Advances in
  Social Networks Analysis and Mining (ASONAM)}, pages 332--339.

\bibitem[Hasheminezhad and Brandes, 2023]{hb23}
Hasheminezhad, R. and Brandes, U. (2023).
\newblock Robustness of preferential-attachment graphs.
\newblock {\em Applied Network Science}, 8.

\bibitem[Hofstad, 2016]{hofstad16}
Hofstad, R. v.~d. (2016).
\newblock {\em Random Graphs and Complex Networks}.
\newblock Cambridge Series in Statistical and Probabilistic Mathematics.
  Cambridge University Press, Cambridge.

\bibitem[Inoue, 2022]{inoue22}
Inoue, M. (2022).
\newblock {\em Modeling and statistical inference of preferential attachment in
  complex networks: Underlying formation of local community structures}.
\newblock PhD thesis, Kyoto University.

\bibitem[Jeong et~al., 2003]{jnb03}
Jeong, H., N\'{e}da, Z., and Barab\'{a}si, A.~L. (2003).
\newblock Measuring preferential attachment in evolving networks.
\newblock {\em Europhysics Letters}, 61(4):567--572.

\bibitem[Kingman, 1993]{kingman93}
Kingman, J. F.~C. (1993).
\newblock {\em Poisson Processes}.
\newblock Oxford University Press.

\bibitem[Krapivsky and Krioukov, 2008]{kk08}
Krapivsky, P. and Krioukov, D. (2008).
\newblock Scale-free networks as preasymptotic regimes of superlinear
  preferential attachment.
\newblock {\em Physical Review E}, 78:026114.

\bibitem[Krapivsky and Redner, 2001]{kr01}
Krapivsky, P.~L. and Redner, S. (2001).
\newblock Organization of growing random networks.
\newblock {\em Physical Review E}, 63(6):066123.

\bibitem[Krapivsky et~al., 2000]{krl00}
Krapivsky, P.~L., Redner, S., and Leyvraz, F. (2000).
\newblock Connectivity of growing random networks.
\newblock {\em Physical Review Letters}, 85:4629.

\bibitem[Larson and Onnela, 2023]{lo23}
Larson, J. and Onnela, J.-P. (2023).
\newblock Maximum likelihood estimation for reversible mechanistic network
  models.
\newblock {\em Physical Review E}, 108.

\bibitem[Lee et~al., 2024]{lef24}
Lee, C., Eastoe, E.~F., and Farrell, A. (2024).
\newblock Degree distributions in networks: Beyond the power law.
\newblock {\em Statistica Neerlandica}, 78(4):702--718.

\bibitem[Li et~al., 2013]{lzcxa13}
Li, H., Zhao, H., Cai, W., Xu, J., and Ai, J. (2013).
\newblock A modular attachment mechanism for software network evolution.
\newblock {\em Physica A}, 392:2025--2037.

\bibitem[Markovich et~al., 2026]{mrv26}
Markovich, N., Ryzhov, M., and Vai\v{c}iulis, M. (2026).
\newblock Inferences for random graphs evolved by clustering attachment.
\newblock {\em Journal of Statistical Planning and Inference}, 241:106332.

\bibitem[{Mohd-{Z}aid}, 2016]{mohdzaid16}
{Mohd-{Z}aid}, M. (2016).
\newblock {\em A Statistical Approach to Characterize and Detect Degradation
  Within the {B}arabasi-{A}lbert Network}.
\newblock PhD thesis, Air Force Institute of Technology.

\bibitem[Newman, 2001]{newman01d}
Newman, M. E.~J. (2001).
\newblock Clustering and preferential attachment in growing networks.
\newblock {\em Physical Review E}, 64:025102.

\bibitem[Overgoor et~al., 2019]{obu19}
Overgoor, J., Benson, A., and Ugander, J. (2019).
\newblock Choosing to grow a graph: Modeling network formation as discrete
  choice.
\newblock In {\em The World Wide Web Conference}, WWW '19, page 1409–1420,
  New York, NY, USA. Association for Computing Machinery.

\bibitem[Pham et~al., 2016]{pss16}
Pham, T., Sheridan, P., and Shimodaira, H. (2016).
\newblock Joint estimation of preferential attachment and node fitness in
  growing complex networks.
\newblock {\em Scientific Reports}, 6.

\bibitem[{P}rice, 1976]{price76}
{P}rice, D. (1976).
\newblock A general theory of bibliometric and other cumulative advantage
  processes.
\newblock {\em Journal of the Assoication for Information Science and
  Technology}, 27(5):292--306.

\bibitem[{R Core Team}, 2025]{cran25}
{R Core Team} (2025).
\newblock {\em R: A Language and Environment for Statistical Computing}.
\newblock R Foundation for Statistical Computing, Vienna, Austria.

\bibitem[Rudas et~al., 2007]{rtv07}
Rudas, A., T\'{o}th, B., and Valk\'{o}, B. (2007).
\newblock Random trees and general branching processes.
\newblock {\em Random Structures and Algorithms}, 31(2):186--202.

\bibitem[Simon, 1955]{simon55}
Simon, H.~A. (1955).
\newblock On a class of skew distribution functions.
\newblock {\em Biometrika}, 42(3-4):425--440.

\bibitem[Voitalov et~al., 2019]{vvvk19}
Voitalov, I., {van der {H}oorn}, P., {van der {H}ofstad}, R., and Krioukov, D.
  (2019).
\newblock Scale-free networks well done.
\newblock {\em Phys. Rev. Res.}, 1:033034.

\bibitem[Wang and Resnick, 2024]{wr24}
Wang, T. and Resnick, S. (2024).
\newblock Random networks with heterogeneous reciprocity.
\newblock {\em Extremes}, 27(1):123--161.

\bibitem[Yule, 1925]{yule25}
Yule, G.~U. (1925).
\newblock {II}.-{A} mathematical theory of evolution, based on the conclusions
  of {D}r. {J}. {C}. {W}illis, {F}. {R}. {S}.
\newblock {\em Philosophical Transactions of the Royal Society of London.
  Series B, Containing Papers of a Biological Character}, 213(402-410):21--87.

\bibitem[Zheng et~al., 2008]{zzlw08}
Zheng, X., Zeng, D., Li, H., and Wang, F. (2008).
\newblock Analyzing open-source software systems as complex networks.
\newblock {\em Physica A}, 387:6190--6200.

\end{thebibliography}

\end{document}